\documentclass[12pt]{article}
\usepackage[dvips]{graphicx}
\usepackage{pdproc}

  \makeatletter 
  \def\@cite#1{[#1]} 
  \makeatother    
\begin{document}

\renewcommand{\thefootnote}{\alph{footnote}}

\title{
 Universal Extra Dimensions and Kaluza-Klein Bound States
}

\author{ Marc Sher}

\address{ 
Particle Theory Group, Department of Physics \\
College of William and Mary, Williamsburg VA 23187 USA
%%%%% You may comment out the e-mail address line below.  
\\ {\rm E-mail: sher@physics.wm.edu}}

\abstract{
We study the bounds states of the Kaluza-Klein (KK) excitations of 
quarks in models of Universal Extra Dimensions.  Such bound states 
may be detected at future lepton colliders in the cross section for 
the pair-production of KK-quarks near threshold.  For typical values 
of model parameters, we find the ``KK-quarkonia'' have widths in the 
10-100 MeV range, and production cross sections of order a few 
picobarns for the lightest resonances.  Two body decays of the 
constituent KK-quarks lead to distinctive experimental signatures.
}

\normalsize\baselineskip=15pt

\section{Bound States}

Two familiar bound states in particle physics are the $J/\Psi$ 
($\overline{c}c$) and the $\Upsilon$ ($\overline{b}b$).   Following 
the discovery of the $\Upsilon$, there was hope that toponium (once 
designated as $\Theta$) would exist and could be studied.  However, 
toponium does not exist.  The bottom quark is so light that there is a 
huge amount of phase space available for top quark decay, and so the 
top quark lifetime is very short.

Thus, the top quark decays before a single orbit can occur.  More 
precisely, $\Gamma_{t}$ is much greater than the toponium level 
splittings, washing out the structure.

In order to have an observable bound state spectrum, the constituent 
width must be less than the onium splittings, i.e. the lifetime of 
the constituent must be long.  For particles with masses in the 
$50-1000$ GeV range, this means that small mass splittings are 
necessary.

This occurs naturally in Universal Extra Dimensions in which the mass 
splittings between the $n=1$ Kaluza-Klein (KK) modes are small.  In this 
talk, which is based on work done with Chris Carone, Justin Conroy and Ismail 
Turan\cite{ccst}, we study bound states of KK-quarks in these models.

\section{Universal Extra Dimensions}

In models with compact extra dimensions, fields which can 
propagate in the extra dimensions have Kaluza-Klein excitations.  
These excitations can cause serious difficulties for electroweak 
radiative corrections, forcing the scale of the extra dimenions to be 
well in excess of a TeV.   However, in Universal Extra 
Dimensions(UED)\cite{ued}, {\it all} fields propagate in the bulk, and 
conservation of KK-number prevents mixing between the $n=1$ modes and 
zero-modes, substantially weakening these bounds down to the collider 
bounds on $1/R$ of approximately $300$ GeV.

Thus, in UED, all fields are in the bulk, KK-number is conserved, 
there are no tree level contributions to electroweak radiative 
corrections, and $1/R$ can be as low as $300$ GeV.  In the absence of 
radiative corrections and electroweak symmetry breaking, all $KK$ 
modes at a given level would be exactly degenerate, with masses given 
by $n/R$.  Electroweak breaking changes these masses to 
$m^{2}=m^{2}_{o} + 1/R^{2}$, but since the $n=0$ mass is negligible for 
all quarks but the top quark, the degeneracy among the other five 
KK-quarks and the KK-photon is not broken.  

Radiative corrections lift this degeneracy, giving the small mass 
splittings needed for bound state formation.  They were calculated by 
Cheng, Matchev and Schmaltz\cite{cms}.   They found that the $n=1$ 
Weinberg angle is very small, and gave expressions for the mass 
splittings under the assumption that the finite parts of counterterms 
vanish at the cutoff scale $\Lambda$.  We will adopt this assumption 
and present results for $\Lambda R = 5, 10, 20$--relaxing it will not 
qualitatively affect our results.  The reader is referred to Ref. 
\cite{cms} for the detailed expressions for the mass splittings.

\section{Decay Modes and Decay Widths of KK-quarks}

Since the mass splittings between the n=1 KK-modes are relatively 
small, and the lightest is absolutely stable, the KK-quarks will be 
relatively long-lived, and can thus form bound states.  In this 
section, we determine the decay modes and widths of the KK-quarks.  For 
definitiveness, we fix $1/R$ to be $500$ GeV and $\Lambda R = 20$; 
the qualitative nature of the results will not be substantially 
affected by this choice.

\subsection{Isosinglet KK-quarks (except the KK-top)}

The masses of the isosinglet KK-quarks (except the KK-top) 
are $572$ GeV.  They can't decay into $KK-W$'s (since they 
are isosinglets) and the decay into $KK-Z$'s is suppresed by the $n=1$ 
Weinberg angle, which is small.  The dominant decay is then
\begin{equation}
    q^{(1)} \rightarrow q^{(0)} \gamma^{(1)}
    \end{equation}
This gives a monochromatic quark with an energy of $67$ GeV, a 
KK-photon which gives missing energy, and the width (for $Q=-1/3$ 
quarks) is within a factor of two of 10 MeV.  Complete expressions for 
the widths, as well as plots as a function of the parameters, 
can be found in Ref. \cite{ccst}. The width for the KK-up 
and KK-charm quarks is a factor of four larger.  The signature for 
pair-production would then be dramatic--two $67$ GeV quarks and 
missing energy.

\subsection{Isodoublet KK-quarks}

Although the phase space is somewhat larger for decay into 
$\gamma^{(1)}$, the larger coupling makes decays into $W^{(1)}, 
Z^{(1)}$ dominant.

The widths are substantially larger (50-100 MeV), but the signatures 
are even more dramatic:
\begin{equation}
    d_{L}^{(1)}\rightarrow u_{L}W^{(1)} \rightarrow u_{L}l\nu^{(1)}
    \end{equation}
    \begin{equation}
	d_{L}^{(1)}\rightarrow u_{L}W^{(1)}\rightarrow u_{L}l^{(1)}\nu
	\rightarrow u_{L}l\nu\gamma^{(1)} \end{equation}
	giving a monochromatic quark and a monochromatic lepton and missing 
	energy for each of the KK-quarks.
	
The decay into KK-Z-bosons (roughly 1/3 of the decays) are
\begin{equation}
    d_{L}^{(1)}\rightarrow d_{L}Z^{(1)}\rightarrow 
    d_{L}ll^{(1)}\rightarrow d_{L}ll\gamma^{(1)}\end{equation}
For all of these decays, all of the final state fermions are 
monochromatic, leading to very distinctive signatures.

\subsection{Isosinglet KK-top}

Since there is no KK-W coupling, flavor changing decays aren't 
possible, and flavor-conserving aren't kinematically allowed.  So 
there are no two-body decays.  The dominant decay will be into a 
KK-$\gamma$ and a virtual top, which then decays into a bottom quark 
and a virtual W which then decays.  The four-body decay is very 
suppressed by phase space, giving a width of about a keV.  This is so 
small that annihilation of the KK-quarkonia (which is tens of keV) 
will dominate. There will be no missing energy.

\section{Production Cross Sections and Resonance Splittings}

The production cross section for a vector resonance can be calculated 
in the standard manner.  For the wavefuntion at the origin, we use 
single gluon exchange, which is expected to be accurate at these high 
energy scales.  Typical cross sections are in the 1-10 picobarn range.

For the splitting between adjacent resonances, we solve the 
non-relativistic radial Schrodinger equation with a single gluon plus 
a linear potential.  Only the first three levels are relevant (recall 
that hydrogen-like spectra level spacings decrease as $1/n^{3}$, so 
the spacings become small very quickly).

\section{Results}

Putting this all together, we find the cross section for isosinglet 
KK-quark production shown in Figure 1.  Note that the signature 
consists of two monochromatic quarks plus missing energy, which is 
quite distinctive with very little background.   For isodoublet 
quarks, the resonances are wider, although the signatures (which 
involve monochromatic leptons as well) are even more distinctive (see 
Ref. \cite{ccst} for a complete set of figures).    Although 
presented for $1/R$ of $500$ GeV, the results\cite{ccst} are very 
similar for $1/R$ as low as $300$ GeV, which would be easily in reach 
of the next linear collider.   

In the figures, we have, of course, neglected beam resolution.  For a 
muon collider, this would be negligible, but it is not negligible for 
an electron collider.  Although the beamstrahlung and initial state 
radiation can cause a substantial spread, their spectra are 
well-know, and it is expected\cite{res} that the mass resolution after 
deconvolution will be as low as $50$ MeV.  This would easily allow the 
resonance structure to be detected.

\begin{figure}[htb]
\begin{center}
\includegraphics*[width=10cm]{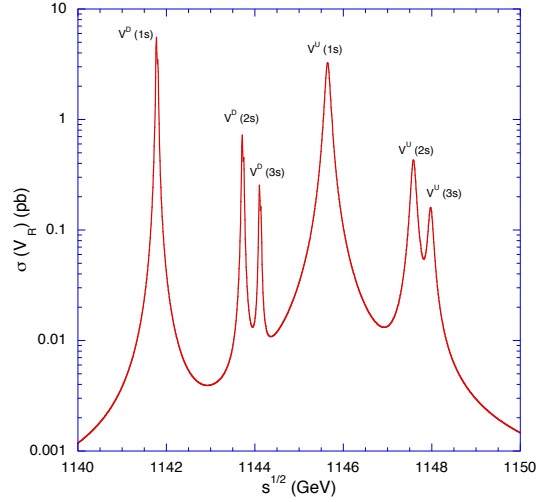}
\caption{%
The cross section for KK-quarkonia formed by isosinglet KK-quarks as 
a function of $\sqrt{s}$ for $1/R=500$ GeV and $\Lambda R=20$.  The 
labels $V^{D}$ refer to the bound states of KK-down, KK-strange and 
KK-bottom quarks, while $V^{U}$ refers to KK-up and KK-charm quarks.
}
\label{fig1}
\end{center}
\end{figure}

\section{Acknowledgements}

I thank my collaborators, Chris Carone, Justin Conroy and Ismail 
Turan.  This work was supported by the National Science Foundation 
under Grant No. PHY-0243400.

\bibliographystyle{plain}

\end{document}